\documentclass[twocolumn,showpacs,preprintnumbers,amsmath,amssymb]{revtex4}

 \usepackage{graphicx}
 \usepackage{amsmath}
 \usepackage{amsfonts}
 \usepackage{amssymb}
 \usepackage{pstricks}
 \usepackage{psfrag}
 \usepackage{epsfig}
 \usepackage[ansinew]{inputenc}

 \newcommand{\vecd}[1]{\mathbf{#1}}
 
 \newcommand{\tens}[1]{\sf {#1}}
 \newcommand{\mat}[1]{\tens{#1}}

\begin{document}

\title{Dumbbell transport and deflection in a spatially periodic potential}

\author{Jochen Bammert \thanks{\email{jochen.bammert@uni-bayreuth.de}} and Walter Zimmermann\thanks{\email{walter.zimmermann@uni-bayreuth.de}}}

\affiliation{Theoretische Physik I, Universit\"at Bayreuth, D-95440 Bayreuth, Germany}

\date{January 9, 2009}

\begin{abstract}
We present theoretical results on the deterministic and stochastic motion of a dumbbell carried by a uniform flow through a three-dimensional spatially periodic potential.
Depending on parameters like the flow velocity, there are two different kinds of movement: transport along a potential valley and stair-like motion oblique to the potential trenches. The crossover between these two regimes, as well as the deflection angle, depends on the size of the dumbbell. Moreover, thermal fluctuations cause a resonance-like variation in the deflection angle as a function of the dumbbell extension.
\end{abstract}

\pacs{47.61.-k, 05.40.-a, 05.60.-k} 

\maketitle

{\it{Introduction.-}}\label{sec: intro} 
The exploration of particle motion in microfluidic devices and energy landscapes has attracted considerable
attention recently, in particular due to its great potential for applications in chemistry and biotechnology
\cite{Grier:2002.1,Vulpiani:2002.1,Dholakia:2003.1,Davis:2006.1,Dholakia:2007.1,Kaoui:2008.1,Steinberg:2006.4}.
With the successful generation of optically induced potentials by holographic laser tweezers \cite{Grier:2002.1,MacDonald:2002.1}, a powerful tool has been introduced for investigations of the classical transport of colloidal particles through those two- or three-dimensional landscapes.
Since the interaction between the colloids and the potential depends on the size and shape
of the particles, such potentials  may also be used in microfluidic 
devices for a spatial decomposition of different particle species.
For instance, spherical particles, carried by a flow with low Reynolds number, can be deflected into a direction oblique to the flow lines, while transversing 
a spatially periodic energy landscape \cite{Dholakia:2003.1,Grier:2004.1}. 
Since the deflection angle is a function of the size and refraction index of 
the particles this behavior may be used for particle sorting.
For extended and deformable objects like vesicles, instead of spherical particles, 
a similar cross-streamline deflection has been found in Poiseuille flow 
even in the absence of a potential \cite{Kaoui:2008.1}.

In this context, interesting questions arise: How are deformable
objects, like dumbbells that are carried by a flow, 
deflected by a spatially periodic potential? What effects are related to the
interplay between the wavelength of the potential and the size of the dumbbell \cite{Bammert:2008.1}? 
What is the role of the hydrodynamic interaction between
the two connected spherical particles?

Our numerical investigation of the flow driven dumbbell motion in spatially periodic potentials can serve as a simple model for other non-spherical objects, such
as dimers \cite{Blaaderen:2005.1}, pom-pom polymers \cite{McLeish:98.1}, or two small beads
connected by polymers. We find in this system
two flow velocity thresholds separating three different regimes
of motion: a locked stage, transport along
a potential trench and a stair-like motion. In addition
we detect a remarkable dependence of the dumbbell deflection angle
 on the ratio $\lambda/b$, where $\lambda$ is the 
wavelength of the periodic potential and $b$ is the length of the
dumbbell. Furthermore, in the presence of stochastic thermal forces, interesting
deflection resonances have been found as a function of
$\lambda/b$.

{\it{Model.-}}\label{sec: numeric} 
We investigate the dynamics of a dumbbell that consists of two beads connected by a linear spring, and driven by the uniform flow $\vecd{u}=u(\cos \alpha, \sin \alpha,0)$ through a three-dimensional spatially periodic potential.
The over-damped motion of the dumbbell is described by the Langevin equation,
\begin{equation}
\label{dgl}
 \dot{\vecd{r}}_i =\vecd{u} + {\mat{H}}_{ij} \left( {\vecd{F}}^{\Phi}_j + \vecd{F}^V_j \right) + \vecd{F}^S_i\,,\qquad (i,j=1,2) \vspace{-2mm}
\end{equation}
for the two bead positions $\vecd{r}_{1,2}$. The spring force, $\vecd{F}^{\Phi}_i=-\nabla\Phi(\vecd{r}_1-\vecd{r}_2)$, between the beads is derived from the harmonic potential,
\begin{equation}
\Phi(\vecd{r}_1-\vecd{r}_2) = \frac{k}{2}\, \left(b-| \vecd{r}_1 -\vecd{r}_2|\right)^2\,, \vspace{-2mm}
\end{equation}
with the spring constant $k$ and the equilibrium bead distance $b$. The force $\vecd{F}_i^V=-\nabla V(\vecd{r}_i)$ is caused by the periodic potential,
\begin{equation}
\label{perpot}
  V(\vecd{r}_i) = -V_0 \left[ \cos\left(qx_i\right) + \cos\left(qy_i\right)+ \cos\left(qz_i\right) \right]\,, \vspace{-1mm}
\end{equation}
with wavelength $\lambda=2\pi/q$ in all three spatial directions, and amplitude $V_0$. The third contribution on the right hand side of Eq. (\ref{perpot}) is included in order to keep the dumbbell axis in the $xy$ plane. In experiments such potentials can be realized by laser tweezers~\cite{Dholakia:2003.1,Grier:2004.1}, where $V_0$ may be changed by varying the intensity of the laser beam.

Eq.~(\ref{dgl}) is a nonlinear function of the bead distance $r_{ij}$ due to the mobility matrix ${\mat{H}}_{ij}$, which describes the hydrodynamic interactions (HI) between the two beads.
Within the Rotne-Prager approximation~\cite{RotnePrager:1969} it has, for $i\neq j$, the form
\begin{align}
 \hspace{-1mm}&{\mat{H}}_{ij}= \frac{3a}{4\zeta r_{ij}} \left[ \left(1+\frac{2}{3}\frac{a^2}{r_{ij}^2} \right){\mat{I}} + \left( 1 - 2 \frac{a^2}{r_{ij}^2} \right) \hat{\vecd{r}}_{ij} \hat{\vecd{r}}_{ij}^T \right] \,,
 \end{align}
with $\vecd{r}_{ij}=\vecd{r}_i-\vecd{r}_j$ and $r_{ij}=|{\bf r}_{ij}|$.
The diagonal component of the mobility tensor, ${\mat{H}}_{ii}=\frac{1}{\zeta}\,\mat{I}$, with unity matrix $\mat{I}$, is inversely proportional to the Stokes friction coefficient, $\zeta=6 \pi \eta a$, of a sphere with effective hydrodynamic radius $a$ within a solvent of viscosity $\eta$.

The uncorrelated stochastic force $\vecd{F}^S_i$ in Eq. (\ref{dgl}) has a vanishing mean value.
The amplitude of $\vecd{F}^S_i$ is determined by the fluctuation dissipation theorem \cite{Dhont:96} and includes the thermal energy $k_BT$:
\begin{align}
 \langle \vecd{F}^S_i(t) \rangle &= 0\,, \\
 \langle \vecd{F}^S_i(t)  \vecd{F}^S_j(t') \rangle &= 2k_BT {\mat{H}}_{ij} \delta(t-t')  \,.
\end{align}
Eq.~(\ref{dgl}) has been solved numerically by an Euler method, as approximate analytical solutions can only be obtained in some special cases.
In our simulations we have fixed the values of $\lambda=1$ and $V_0=1$, while all other lengths and energies are given in terms of them. Furthermore, we choose a radius of $a=0.2$ for the beads, a spring constant of $k=10$ and a Stokes friction coefficient of $\zeta=10$.
Angles are given in radians with respect to the $x$ axis. The time scale is set by the relaxation time of the spring $\tau = \zeta/k$, therefore the velocities are given in units of $\lambda /\tau$.
Since the flow direction $\vecd{u}$ is restricted to the $xy$ plane, the periodic variation of the potential in the $z$ direction keeps the dumbbell axis in a single minimum with respect to $z$, i.e. $z=0$.
At the initial state of the calculations, we align the dumbbell parallel to the flow lines.

\begin{figure}[ht]
\vspace{-4mm}
  \begin{center}
\includegraphics[width=0.95\columnwidth]{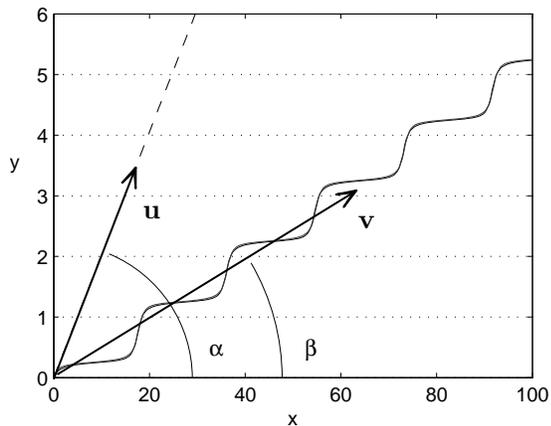}
  \end{center}
\vspace{-5mm}
  \caption{A sample deterministic trajectory of the center of mass of a dumbbell (solid line) in the $xy$ plane, which is driven by the flow $\vecd{u}$ through a three-dimensional periodic potential $V$. The mean velocity of the dumbbell is denoted by $\vecd{v}$ and the horizontal dotted lines indicate the valleys of $V$ parallel to the $x$ axis.
}
\label{traj}
\end{figure}

A typical situation investigated in this work is sketched in Fig.~\ref{traj}. The dotted lines in the figure indicate the valleys of the periodic potential that are parallel to the $x$ axis (trenches parallel to the $y$ axis are not shown). The streamlines of the flow field $\vecd{u}$ cross these valleys with an angle $\alpha<\pi/4$. This flow imposes a drag force $\vecd{F}^{\zeta}$ on the beads of the dumbbell, which is proportional to $u$, and which depends on the bead distance $r_{12}$. If $\vecd{F}^{\zeta}$ is large enough the dumbbell moves with the mean velocity $\vecd{v}$. Its deflection angle is denoted by $\beta$.

\begin{figure}[htb]
\begin{center}
 \includegraphics[width=0.9\columnwidth]{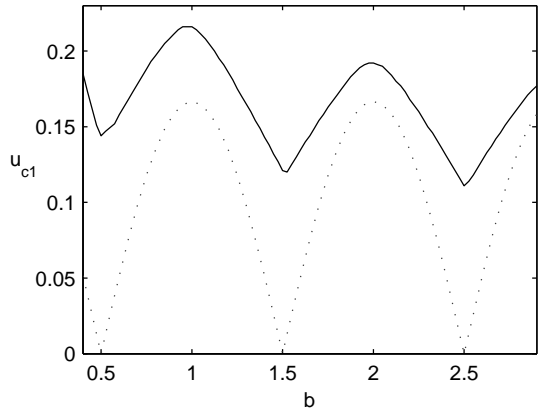}
\end{center}
 \vspace{-5mm}
  \caption{The threshold velocity $u_{c1}$ for driving the dumbbell along the $x$ direction is shown as a function of the equilibrium bead distance $b$ (solid line). The dotted line indicates the analytical result calculated from Eq.~(\ref{potforce}).
}
\label{u0c_th}
\end{figure}

{\it{Results.-}}\label{sec: results} 
We first consider the dumbbell motion in the limit of vanishing thermal noise. For small flow strengths $u$  the dumbbell is captured by the potential. If the flow velocity ${\bf u}=u(1,0,0)$ exceeds a certain threshold value $u_{c1}$, the dumbbell surmounts the saddles of the potential at a height $V_0$ and moves along a wavy potential trench with a periodically changing velocity parallel to the $x$ axis. The threshold velocity $u_{c1}$ is a function of the bead distance $b$ as shown in Fig.~\ref{u0c_th} for one parameter set.

The dumbbell aligned with its axis parallel to the $x$ direction and with a fixed bead distance $b$ experiences a potential force related to Eq.~(\ref{perpot}),
\begin{align}
\label{potforce}
 F^V(x,x+b)=-\frac{\partial}{\partial x}\left(V(x)+V(x+b)\right)\,. \vspace{-2mm}
\end{align}
This force vanishes at bead distances $b =(2n+1) \lambda/ 2$, and accordingly the required flow velocity for driving the dumbbell continously through the potential also vanishes. However, at finite values of the spring constant $k$, corresponding to flexible objects like polymers etc., the dumbbell is stretched and compressed during its motion through the spatially modulated potential.
With this degree of freedom the threshold flow velocity, $u_{c1}$, still has pronounced minima at the distances $b=(2n+1) \lambda /2$ but it does not vanish as in the case of a fixed bead distance, especially when the HI is taken into account. This can be seen in Fig.~\ref{u0c_th}, where the dotted line indicates the threshold velocity of a rigid dumbbell. Moreover the HI reduces the drag force on the two beads. Since this reduction is proportional to $1/r_{12}$, the maxima of the curve $u_{c1}(b)$ decrease with increasing values of $b$.

\begin{figure}[ht]
\vspace{-4mm}
  \begin{center}
\includegraphics[width=0.9\columnwidth]{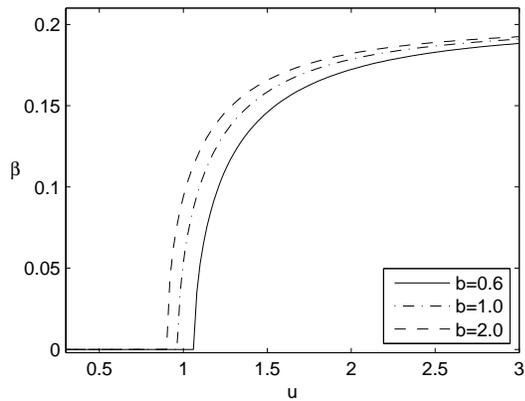}
  \end{center}
\vspace{-5mm}
  \caption{The deflection angle $\beta$ of the dumbbell is shown as a function of the 
 flow velocity $u$ at $\alpha=0.2$ for different values of the bead distance
 $b$. In the regime of $\beta=0$ the dumbbell moves along a potential trench.
}
\label{ud}
\end{figure}

If the flow velocity, $\vecd{u}= u(\cos \alpha, \sin \alpha,0)$, is increased further, one reaches, for finite angles $\alpha$ and beyond a second threshold velocity $u_{c2}>u_{c1}$, another regime of dumbbell motion. If the $y$ component of the drag force $\vecd{F}^\zeta$ is large enough, the beads frequently jump to a neighboring trench of the potential parallel to the $x$ axis. Consequently the dumbbell performs a stair-like motion in the $xy$ plane as indicated by the solid line in Fig.~\ref{traj}.
The mean direction $\vecd{v}$ of this motion has a finite deflection angle $\beta$, which depends on $u$ as shown in Fig.~\ref{ud}. In the limit of large values of $u$, $\beta$ approaches the inclination angle $\alpha$ of the flow $\vecd{u}$ as expected.

\begin{figure}[ht]
  \begin{center}
\includegraphics[width=0.9\columnwidth]{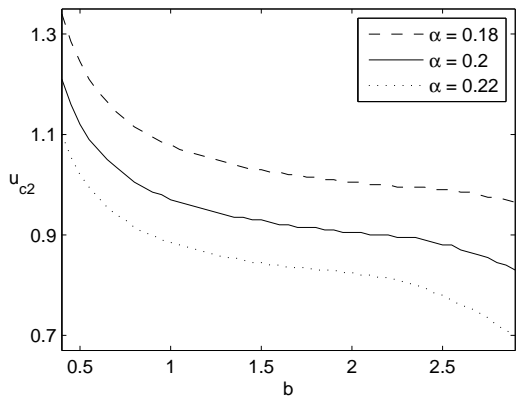}
  \end{center}
\vspace{-5mm}
  \caption{The critical velocity $u_{c2}$ for the onset of the stair-like motion is shown as function of $b$ for three different inclination angles $\alpha$ and $u=1$. }
\label{u0c2}
\end{figure}

Since the drag force imposed by the flow on the dumbbell depends on the distance between the two beads, $\vecd{r}_{12}$, the deflection angle is sensitive to the equilibrium length of the spring as indicated for three different values of $b$ in Fig.~\ref{ud}.
Consequently the critical value $u_{c2}$, which marks the transition to finite values of $\beta$, is also sensitive to the bead separation.
This monotone $b$-dependence of $u_{c2}$ is shown for three different angles $\alpha$ in Fig.~\ref{u0c2}.

\begin{figure}[ht]
  \begin{center}
\includegraphics[width=0.9\columnwidth]{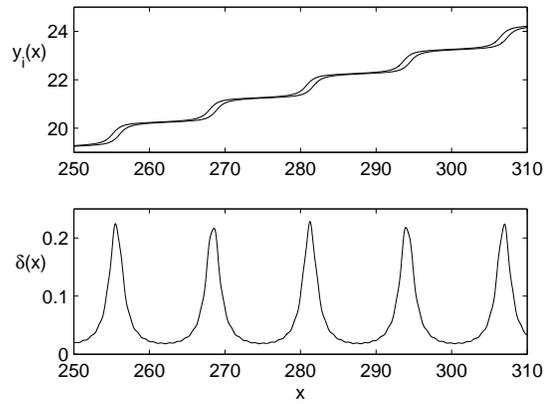}
  \end{center}
\vspace{-5mm}
  \caption{The trajectories of the two beads in the $xy$ plane are shown in the upper panel for the parameters $\alpha=0.2, u=1, b=1.5$. The lower panel shows the corresponding evolution of the orientation angle $\delta$ of the dumbbell.
 }
\label{deltat}
\end{figure}

During its stair-like motion in the $xy$ plane the dumbbell changes its orientation, which is described by the angle $\delta$ between $\vecd{r}_{12}$ and the $x$ axis, while passing the saddles of the potential  as indicated by the lower panel in Fig.~\ref{deltat}.
 The amplitude of $\delta$ depends on the dumbbell length $b$.
With increasing values of $u$ and $\alpha$ those reorientation steps occur more often, accordingly the typical frequency of $\delta(t)$ increases as well as the angle $\beta$.

\begin{figure}[ht]
  \begin{center}
\includegraphics[width=0.9\columnwidth]{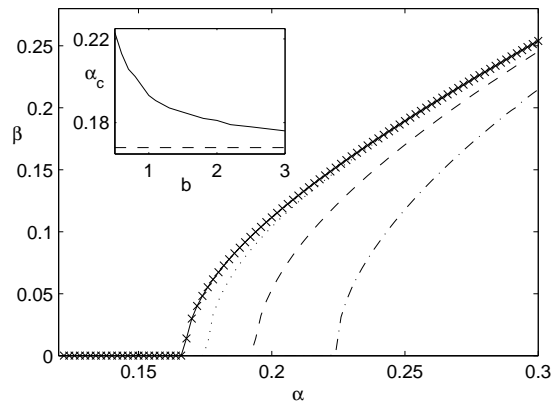}
  \end{center}
\vspace{-5mm}
  \caption{The deflection angle $\beta$ is shown as a function of the flow inclination angle $\alpha$ for different values of the dumbbell-extension (dotted: $b=3$, dashed: $b=1$, dash-dotted: $b=0.5$) and $u=1$. The solid line is obtained for a dumbbell in the free-draining limit (without HI) and does not change with $b$. It coincides with the curve $\beta(\alpha)$ of a single bead (crosses). The inset shows the critical angle $\alpha_c$ versus $b$ for a dumbbell with HI (solid) and without HI (dashed).}
\label{ad}
\end{figure}

Alternatively one may vary the inclination angle $\alpha$ instead of the flow velocity $u$.
If $u$ is fixed at values larger than $u_{c1}$, the drag force $\vecd{F}^\zeta$ is too weak at a small angle $\alpha$ to move the dumbbell across the potential barriers in the $y$ direction.
But above a critical value $\alpha_c$, which again depends on the length $b$ of the dumbbell, one finds the transition from the motion along a potential valley to stair-like trajectories. The relation between $\beta$ and the inclination angle $\alpha$ is shown in Fig.~\ref{ad}.
Note, if the HI is neglected, all curves $\beta(\alpha)$ for dumbbells of different sizes coincide with the one of a single bead \cite{Grier:2004.2}, which means that without HI the deflection angle has no $b$-dependence.

Just below the critical values $\alpha_c$ and $u_{c2}$ thermal fluctuations support the passage of the beads between neighboring potential trenches and therefore may induce a stair-like motion in this parameter range. We investigated this influence of the thermal noise by taking into account the additive stochastic force $\vecd{F}^S_i$ in the equation of motion (\ref{dgl}).

\begin{figure}
  \begin{center}
\includegraphics[width=0.9\columnwidth]{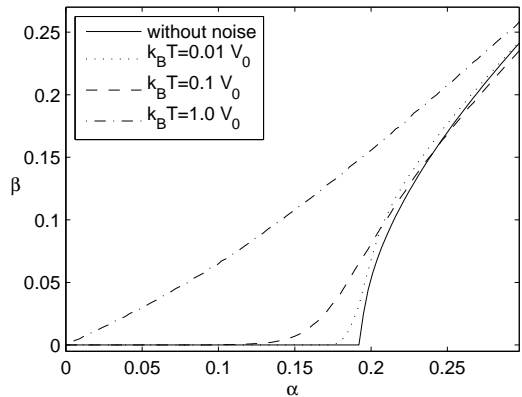}
  \end{center}
\vspace{-5mm}
  \caption{The deflection angle $\beta$, averaged over
$500$ independent runs, of a dumbbell with $b=1$ as a function of the flow angle $\alpha$ at $u=1$ for different noise amplitudes.}
\label{adTno}
\end{figure}

The deflection angle $\beta$ as a function of $\alpha$ is shown for different values of the noise amplitude in Fig.~\ref{adTno}. These numerical results, which are obtained by averaging over $500$ independent runs, show that the dumbbell motion is most sensitive to thermal fluctuations close to the critical angle $\alpha_c$ or close to the critical velocity $u_{c2}$. The onset of the stair-like motion is shifted to smaller values of $u$ or $\alpha$ as explained above, and in the limit of small noise amplitudes the deterministic result is approached.
In the case of a noise amplitude in the order of $V_0$, the angles $\alpha$ and $\beta$ become similar, indicating that the thermal motion is strong enough to kick the beads across  the saddles of the potential. In this case all dumbbells are equally deflected and particle sorting is no longer possible.

\begin{figure}
  \begin{center}
\includegraphics[width=0.9\columnwidth]{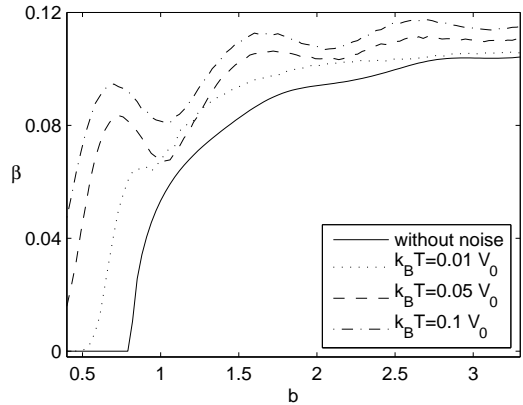}
  \end{center}
\vspace{-5mm}
  \caption{The deflection angle $\beta$, averaged over
$500$ independent runs, shown as function 
of the dumbbell size $b$. It exhibits an interesting resonance like behavior, especially 
for the noise amplitude $k_BT/V_0=0.1$.
Parameters: $u=1$, $\alpha=0.2$.}
\label{bdT}
\end{figure}

Another interesting dependence of the deflection angle on the noise strength is shown in Fig.~\ref{bdT}, where $\beta$ exhibits an interesting resonance like behavior as a function of the bead distance $b$. The deflection angle is clearly reduced in the regime, where the dumbbell length is a multiple of the wavelength and a higher excitation energy of the dumbbell is required.
According to the behavior shown in Fig.~\ref{bdT}, dumbbell sorting with respect to the size is most efficient within the range $0.5 < b/\lambda< 1.5$ in the limit of $k_BT \ll V_0$.

{\it{Conclusions and discussion.-}}\label{conclusion}
We investigated the flow induced transport of mesoscopic and deformable dumbbells through a spatially periodic potential and found three different regimes of motion.
The transitions between these regimes depend sensitively on the flow velocity, on the size of the dumbbell, and therefore on the hydrodynamic interactions between the two beads.
Below a first critical velocity the dumbbell is locked by the potential while above this threshold the two beads start to move along a potential valley. The critical flow strength depends on the length of the dumbbell in a periodic way.
If the flow velocity increases further at a finite inclination angle and crosses a second threshold, then a transition to a stair-like motion occurs. The preferred mean direction of this motion is neither along the flow nor along a symmetry direction of the potential, and the resulting deflection angle depends on the length of the dumbbell.
This allows for the use of periodic potentials, with their symmetry axis oblique to the flow direction, to deflect particles with respect to their size. The strongest sensitivity of the deflection angle on the dumbbell length, which may be employed for particle sorting purposes, occurs in the cross over regime to the stair-like motion.
This cross over is sensitive  to thermal noise because thermal fluctuations make the transition less sharp.
Moreover they induce a resonance like structure in the deflection angle caused by the interplay between the wavelength of the potential and the size of the dumbbell.
These effects, which weaken the efficency of particle sorting, may be suppressed by increasing the potential amplitude and the flow velocity simultaneously. 
Note, that the results in the regime of the transition to the stair-like motion can be obtained qualitatively too if an experimentally easier accessible potential is chosen by neglecting the first contribution on the right hand side of Eq.~(\ref{perpot}).

We would like to thank L. Holzer and S. Schreiber for instructive discussions.
This work has been supported by the German Science Foundation through
the priority program on micro- and nanofluidics SPP 1164.

\end{document}